\documentclass[usenatbib,twocolumn]{mn2e}
\usepackage{graphicx,color}
\usepackage{amssymb}
\usepackage{subfigure}
\usepackage{multirow}

\newcommand{\beq}{\begin{equation}}
\newcommand{\eeq}{\end{equation}}

\def\beq{\begin{equation}}
\def\ee{\end{equation}}
\def\lsim{\mathrel{\rlap{\lower4pt\hbox{\hskip1pt$\sim$}}
    \raise1pt\hbox{$<$}}}
\def\gsim{\mathrel{\rlap{\lower4pt\hbox{\hskip1pt$\sim$}}
    \raise1pt\hbox{$>$}}}

\title[Starspots versus stellar age]
 {Determination of {the} Starspot Covering Fraction as a function of Stellar Age from Observational Data}

\author [Nichols-Fleming, Blackman]
{Fiona Nichols-Fleming$^{1,2}$ 
\thanks{E-mail: {fiona\_nichols-fleming@brown.edu}},
Eric G. Blackman$^{2,3}$ 
\thanks{E-mail: blackman@pas.rochester.edu}\\ 
$^{1}$Department of Earth, Environmental and Planetary Sciences, Brown University, Providence RI, 02912, USA\\
 $^{2}$Department of Physics and Astronomy, University of Rochester, Rochester NY, 14627, USA\\
$^{3}$Laboratory for Laser Energetics,  University of Rochester, Rochester NY, 14623, USA\\}

\begin{document}

\date{}
\pagerange{\pageref{firstpage}--\pageref{lastpage}} \pubyear{}
\maketitle
\label{firstpage}
\begin{abstract}

The association of starspots with magnetic fields leads to an expectation that quantities which correlate with magnetic field strength may also correlate with {starspot} coverage. Since younger stars spin faster and are more magnetically active, assessing whether {starspot} coverage correlates with shorter rotation periods and stellar youth tests these principles. Here we analyze the {starspot} covering fraction versus stellar age for M{-}, G{-}, K{-}, and F{-}type stars based on previously determined variability and rotation periods of over 30,000 {\textit{Kepler}} main-sequence stars. We determine the correlation between age and variability using single and dual power law best fits. We find that {starspot} coverage does indeed decrease with age. Only when the data {are} binned in an effort to remove the effects of activity cycles of individual stars, do statistically significant power law fits emerge for each stellar type. {Using bin averages,} we then find that the {starspot} covering fraction scales with the {X}-ray to bolometric ratio to the power $\lambda$ with {$0.22\pm 0.03 < \lambda < 0.32\pm 0.09$} for {G-type} stars of rotation period below 15 days and for the full range of F{-} and M{-type} stars.  For K{-}type stars, we find two branches of $\lambda$ separated by variability bins, with the lower branch showing nearly constant starspot coverage and the upper branch
 {$\lambda \sim 0.35\pm 0.04$.}
G{-}type stars with periods longer than $15$ days exhibit   a transition to steeper power law of 
{$ \lambda \sim 2.4 \pm 1.0$.}
The  potential connection to previous  rotation-age measurements suggesting a magnetic breaking transition at the solar age, corresponding to period of $24.5$ is also of interest.
 \end{abstract}


\section{Introduction}

During the Sun's 11-year activity cycle, dark sunspots appear for timescales ranging from days to a few months \citep{Petrovay1997}. The appearance and disappearance of these sunspots produces  brightness variations on different timescales which can be detected in total solar irradiance (TSI) data \citep[e.g.][]{Domingo2009}. Stars other than the Sun also  have activity cycles and variations in chromospheric emission \citep{Wilson1978,Vaughan1980,Noyes1984a,Noyes1984b}. These variations are typically measured with the S index,  which is based on the Ca II H and K lines \citep{Vaughan1978}, but can also be observed with long-term brightness variations \citep{Baliunas1985,Olah2000,Olah2002,Messina2002}.

\cite{Baliunas1995} found a separation in mean values of S indices for fast{-} and slow{-}rotating {main-sequence} stars which have higher and lower S values, respectively, indicating that younger stars exhibit more activity than older stars. Additionally, two distinct branches have been {identified} in the relation between rotation and {dynamo} cycle period, {which have been labeled the active (A) and inactive (I) sequences, where the inactive branch applies to slower rotators and has  rotation to dynamo cycle period ratios about six times larger than the active branch.} \citep{Brandenburg1998b,Saar1999}. There is also  a separate branch for fast rotators with rotation periods less than three days. The A and I branches were confirmed by \cite{Bohm2007} {who also saw the Sun placed between the two branches, but  
evidence for  these multiple branches has not been robustly confirmed by the more recent study of \cite{doNascimento2015}.}

Indicators of stellar activity are believed to be driven by the stellar magnetic dynamo \citep{Parker1955,Wilson1966,Kraft1967,Charbonneau2014} which is in turn thought to be driven by some combination of convection, rotation, and differential rotation \citep{Duvall1984}. {Given this correlation, the dynamo dependence on rotation for slow rotators like the Sun suggests that an overall decrease in indicators of stellar activity over much longer time scales should correlate with spin-down of the stars driven by their stellar winds \citep[e.g.][]{Skumanich1972,Pizzolato2003,Mamajek+2008}. The stellar winds likely depend on energy sources that strengthen with magnetic field and the magnetic dynamo \citep{Cranmer2011}, and therefore these correlations create a positive feedback loop 
\citep[e.g.][]{Blackman2016}}.

Large sample observations allow extraction of empirical scalings of coronal activity versus rotation period or Rossby number {($Ro$), where $Ro$ is} the ratio of rotation period to a stellar model{-}dependent eddy turnover time  \citep{Wright2011,Reiners2014}. \cite{Wright2011} show that for older main-sequence stars with Ro $>0.13$, the ratio of the X-ray to bolometric flux of the star, used as a metric of activity, varies as Ro$^{-\zeta}$ where $2\leq \zeta \leq 3$. For younger main-sequence stars with Ro $<0.13$, the relationship saturates. {\citet{Reiners2014}} emphasizes that the eddy turnover time is itself luminosity dependent and so favors relations directly scaling activity to rotation period. For present purposes, we ignore stars in the saturated regime as the population has multi-valued periods for a given luminosity and because we  exclude  stars  so young that accretion is still influencing their behavior.

Since  starspot coverage and X{-}ray luminosity  are both closely related to the stellar magnetic dynamo, we may expect  a relationship between starspot coverage and rotation period of stars, and therefore between {starspot} coverage and the ratio of X-ray to bolometric flux of the star. \cite{Notsu2019} found that {starspot} area doesn't depend on rotation period for young stars, but begins to decrease quickly for rotation periods greater than about 12 days for Sun-like stars. This observed trend is based on chromospheric lines, so further investigation of the trend based on photometric brightness is timely.

There have been  some previous efforts  to characterize starspot coverage from photometric brightness variations measured by the {\textit{Kepler Space Telescope}} \citep{Savanov2017b, Savanov2017a, Dmitrienko2017, Savanov2018}. These authors measure {starspot} coverage of various samples of main-sequence stars, with an S factor based on changes in the mean radiation flux of the stars. \cite{Savanov2017a} found two groups of solar-type stars, active and inactive, based on the stars' S values. The active group is only about ten percent of the sample, but has a mean S value about the same as that of the Sun. The active sample has a decreasing trend in S value with age while the inactive sample have a constant S value. \cite{Dmitrienko2017} look at the relationship between S values for M dwarfs and Rossby numbers and find that it resembles that found in \cite{Wright2011} with saturation at the same value of $Ro$. Although these works identify a decreasing trend in {starspot} coverage as stars age, further  efforts to identify a  functional form of this relationship are warranted and  to assess whether there is evidence for any dynamo transitions that have been purported for solar{-}type stars in smaller samples \citep{vanSaders2016,Metcalfe+2017}.

In Section 2, we discuss the theoretical connection between stellar age, X-ray luminosity, and rotation. In Section 3, we discuss the observational sample used in this work. In Section 4, we discuss our analysis methods. In Section 5, we present our results for single and dual power law relationships as well as trends in vertically binned data. In {Section 6}, we discuss the theoretical implications of this work. We conclude in {Section 7.}

\section{Connecting stellar activity, age, rotation and starspot coverage}

Different theoretical or semi-theoretical approaches have been pursued 
to understand  aspects of the observed relationship between coronal activity and rotation period \citep{Pallavicini1981, Noyes1984a, Vilhu1984, Micela1985, Hartmann1987, Randich2000, Montesinos2001, Pizzolato2003, Wright2011, Reiners2014, Vidotto2014, Blackman2015}. 
To highlight the role of starspot covering fraction in this effort, note for example {the work of 
\cite{Blackman2015} in which they connected the activity evolution over stellar spin down times
 to saturation of a dynamo model for fast rotators. In doing so, the authors} parameterized a relation between the ratio of X-ray 
to bolometric luminosity ratio $\frac{\mathcal{L}_X}{\mathcal{L}_*}$
and the solid angle that the magnetic field passes through, $\Theta$ of the form {
\beq
{\Theta=\Theta_0\left( {{\mathcal{L}_X}/\mathcal{L}_*\over 6.6\times10^{-7}}\right)^\lambda}
\label{theta},
\label{1}
\eeq
where} $\Theta_0$ is the cycle averaged value of $\Theta$ for the present Sun, and
 $\lambda$ is a needed parameter to account for the decreasing {starspot} coverage fraction with age discussed further below. This work also provides an expression for the ratio of X-ray to bolometric luminosity with  dependence
\beq
\frac{\mathcal{L}_X}{\mathcal{L}_*}\propto
\left(\frac{s^{1/3}}{1+2\pi s\ Ro}\right)^2\Theta, 
\label{bt_19}
\eeq
based on the magnetic energy flux sourced by the internal dynamo 
that buoyantly rises into the corona, and  averaged over a dynamo cycle period. {Here $s$ is a constant that accounts for differential rotation in the models of \citet{Blackman2015} and $Ro$ is the Rossby number.} From Equations (\ref{theta}) and (\ref{bt_19}), 

{\beq
{\frac{\mathcal{L}_X}{\mathcal{L}_*}\propto\left(6.6\times10^{-7}\right)^{-\lambda/1-\lambda} \Theta_0^{1/1-\lambda}\left(\frac{s^{1/3}}{1+2\pi s\ Ro}\right)^{2/(1-\lambda)}}
\eeq
which, when removing the constants from the proportionality, gives}
\beq
{\frac{\mathcal{L}_X}{\mathcal{L}_*}\propto\left(\frac{s^{1/3}}{1+2\pi s\ Ro}\right)^{2/(1-\lambda)}}.
\label{prop}
\eeq

{For $Ro \gg1$  and a chosen value of $\lambda$,  Equation (\ref{prop}) can provide a value for $\zeta$, the parameter describing the relationship between Rossby number and X-ray to bolometric flux from \citet{Wright2011}. If $\lambda=0$, then $\zeta\sim 2$ whereas if $\lambda>\frac{1}{3}$, then $\zeta>3$.} As observations have shown that $2\leq\zeta\leq3$ \citep{Wright2011},
an empirical value of  $0\leq\lambda\leq\frac{1}{3}$ would be expected.
The scaling relation of equation (\ref{prop}) was also used in a subsequent minimalist unified theory for the evolution of X-ray luminosity, rotation, magnetic fields, and mass loss of main-sequence stars in the unsaturated regime  \citep{Blackman2016}. Direct constraints on $\lambda$ are timely not only for use in such holistic theoretical approaches, but also 
as a fundamental observational constraint to be explained from first principles.

\section{Observational Sample}

\cite{McQuillan2014} derived rotation periods and measured the photometric variability of 34,030 main-sequence stars based on three years of {\textit{Kepler}} observations. Rotation periods are found with an autocorrelation method that has been shown to be more robust than the more common Fourier methods \citep{McQuillan2013}. The value of a star's variability is defined as the difference between the 95th and 5th percentile of the normalized flux for the star in a single rotation period,  then  averaged over all rotation periods contained over the duration of the observations.

We don't consider the 2,487 stars in the sample with rotation periods shorter than 3 days. {These stars were removed due to the risk of ongoing accretion affecting variability,} leaving a dataset of 31,543 stars. {When looking at the 2,487 fast rotators, there is indeed a large amount of variability indicating that there is no evidence for a saturated regime in this sample that is not masked by the highly variable regime.} \cite{McQuillan2014} also provide stellar parameters for the sample {from either the \textit{Kepler} Input Catalog (KIC) or \citet{Dressing2013},} including mass and effective temperature. The sample covers masses $0.256 \le M/M_\odot \le 1.28$ and effective temperatures $3204 \le T \le 6499$ K.

\begin{figure*}{
\begin{center}
\includegraphics[width=0.9\textwidth]{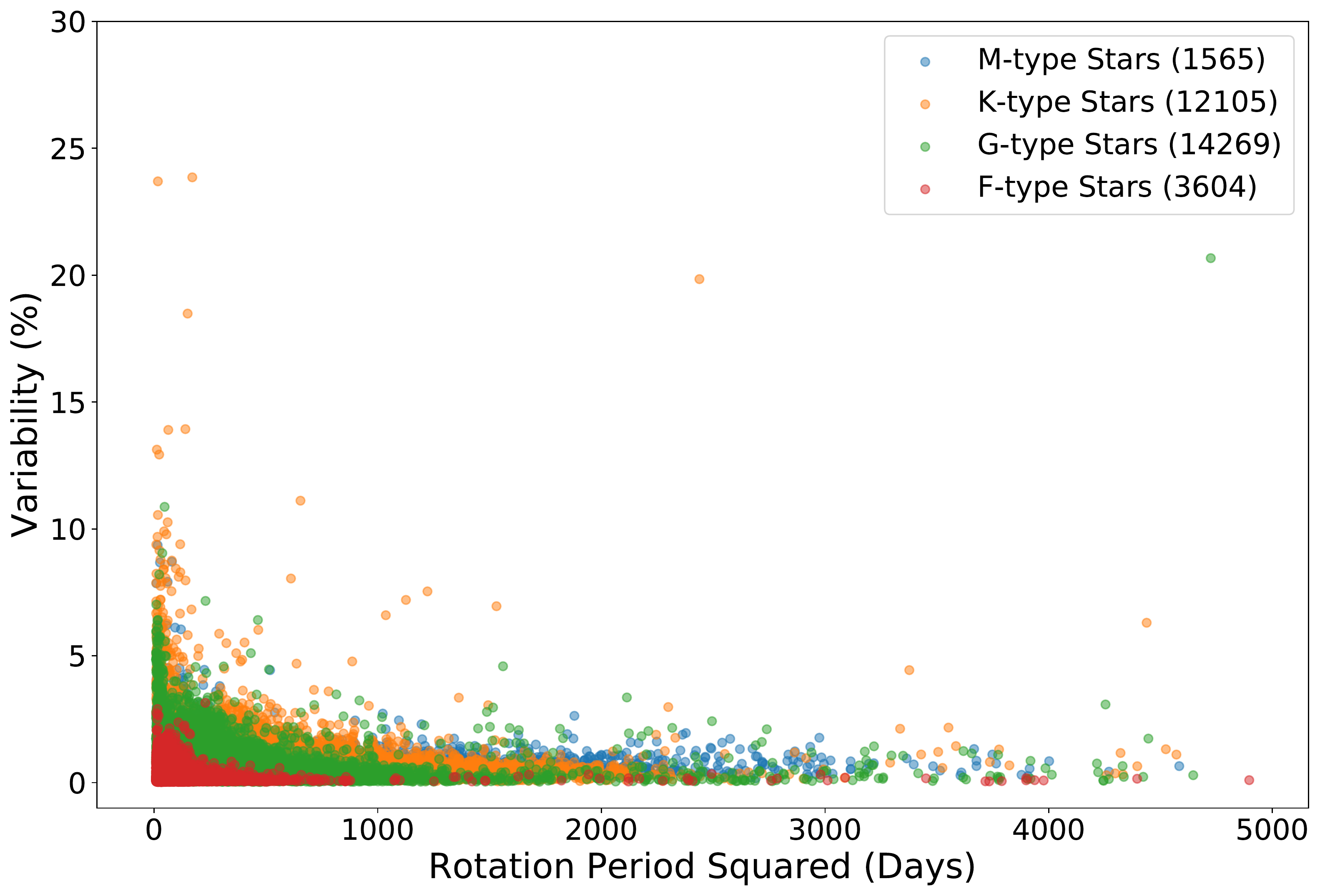}
\end{center}
\caption{Variability versus rotation period squared for the entire sample of 31,543 stars with rotation periods above 3 days from \citet{McQuillan2014}. Rotation period squared is used as a proxy for age where larger $P^2$ is older, so this plot demonstrates the evolution of stellar activity with age for main{-}sequence stars. The sample is divided into stellar types based on their effective temperatures and are shown here in different colors: blue for M-type, orange for K-type, green for G-type, and red for F-type. The number of stars for each stellar type is shown in the upper right corner.}\label{fig:full_sample}}
\end{figure*}

Stars of different  types may evolve differently, so we use the effective temperatures to divide the sample into stellar type subgroups {using the temperature ranges from \citet{Drilling2000}. From this we find that we have 3,604 F-type stars ($5940 < T \leq 7300$), 14,269 G-type stars ($5150 < T \leq 5940$), 12,105 K-type stars ($3840 < T \leq 5150$), and 1565 M-type stars ($3170 < T \leq 3840$). The variability versus rotation period  squared ($P^2$) for the entire sample is shown in Figure \ref{fig:full_sample} with colors indicating each star's spectral type and it can be seen that all spectral types have decreasing variability, interpreted as spot coverage, with increasing rotation period (or age).} Although the G{-} and K{-type} subgroup sizes are the largest, even the samples of F{-} and M{-}type stars are still much larger than those of most previous works, and comparable to those of \cite{Savanov2017b, Savanov2017a, Dmitrienko2017, Savanov2018}.

\section{Methods}

In this work we assume that the {starspot} coverage fraction is directly proportional to the average variability of the star. {This may not be valid for all variability regimes, in part because stars may have a certain amount of continuous starspot coverage. But this is  presently hard to account for, so we proceed with this assumption of proportionality for now.}
 \cite{Blackman2015} explain that the fraction of the solid angle that the field rises through on its journey to the corona is plausibly proportional to the areal fraction of starspots and therefore stellar variability can give a measure of the solid angle through which the field rises. Skumanich's {``}law" states that the rotation period of a star is approximately proportional to the square root of its age \citep{Skumanich1972}, so we look for a relationship between the variability of the star and its rotation period squared to represent the {starspot} coverage evolution in time. To quantify this relationship between {starspot} coverage fraction and age, {we determine the correlation of these quantities and then attempt to fit the observations with  single and  dual power laws.}

Specifically, we determine the correlation of the data using a Pearson correlation coefficient on the logarithmic relationship between the rotation period squared and the average variability the stars. The Pearson correlation coefficient is defined as 
{\beq
r_{P,V}=\frac{cov(P,V)}{\sigma_P\sigma_V},
\eeq
where $cov(P,V)$ is the covariance of $P^2$ and the variability $V$, and $\sigma_P$ and $\sigma_V$ are the standard deviations of $P^2$ and the variability $V$, respectively.}

We also quantify the statistical significance of our best fit functions using a reduced chi square value. The reduced chi square is given by 
{\beq
\chi^2=\frac{1}{N-n}\sum_i^N\frac{\left(V_i-f(P_i)\right)^2}{\sigma_i^2},
\label{chi}
\eeq
where} $N$ is the number of data points, $n$ is the number of fit parameters, $V_i$ is the variability value from the data for a {given $P^2$, $f(P_i)$ is the best fit value for a given $P^2$, and $\sigma_i$ is the variance at a given $P^2$.} To be statistically significant, this reduced $\chi^2$ value should be close to one.

As another way to quantify the change in {starspot} covering fraction with age, we determine values of the  free parameter $\lambda$ used in the holistic  model of stellar activity evolution of \cite{Blackman2016} for each stellar type subsample. From equation (\ref{1}), 
 we find the following expression for $\lambda$.
\beq
\lambda=\frac{\frac{d}{dt}\ell n\left(\Theta\right)}{\frac{d}{dt}\ell n\left(\mathcal{L}_X/\mathcal{L}_*\right)}.
\label{lam}
\eeq

To determine an expression for the ratio of X-ray to bolometric luminosity of a star as a function of its age $t$, we use the results from \cite{Wright2011}:
{\beq
\frac{\mathcal{L}_x}{\mathcal{L}_*}=C*Ro^\beta
\eeq
where $\beta\equiv -\zeta=-2.70\pm0.13$,} in combination with Skumanich's law.  The Rossby number is proportional to the rotation period divided by the eddy turnover time, and the age is proportional to the rotation period squared. Thus if we assume that the eddy turnover time is constant for a particular stellar type, we find that $Ro\propto P\propto t^{1/2}$ and obtain 
{\beq
\frac{\mathcal{L}_x}{\mathcal{L}_*}=C*t^{\alpha},
\label{Lage}
\eeq
where $\alpha=\frac{1}{2}\beta=-1.35\pm0.065$. Equation (\ref{Lage}) and the assumption that $\Theta \propto Variability \propto (P^2)^b$ where $b$ is the derived power of our fit, allows us to write our expression for $\lambda$ (equation \ref{lam}) in the simpler form}\footnote{An alternative approach is  to use the results of \cite{Reiners2014} where the ratio of X-ray to bolometric luminosity is a function of rotation period and radius rather than the Rossby number.
Presently we  restrict ourself  to using the expression from \cite{Wright2011}.}
\beq
\lambda = \frac{b}{\alpha}.
\label{ourlam}
\eeq

To find the error in our values of $\lambda$, we use the basic error propagation formula
\beq
\sigma_f^2=\sum_{i=1}^N \left(\frac{df}{dx_i}\right)^2 \sigma_{xi}^2,
\label{errprop}
\eeq
which reduces to  
\beq
\sigma_\lambda^2 = \left(\frac{b}{\alpha^2}\right)^2 \sigma_\alpha^2 + \left(\frac{1}{\alpha}\right)^2  \sigma_b^2,
\label{ourerr}
\eeq
where $\sigma_\alpha$ is the derived error in $\alpha$ from \cite{Wright2011} and $\sigma_b$ is the derived error in the power of an individual best fit power law found in this work. 

For the dual power law fits, we again look at the Pearson correlation coefficients, reduced $\chi^2$ values, and values of $\lambda$ for the different age regimes. We look at the Pearson $r$ value for all stars with periods longer and shorter than {a chosen transition period and define the formula}
\beq
F_{T} \equiv 1-abs(r_{low})\frac{n_{low}}{N}-abs(r_{high})\left(1-\frac{n_{low}}{N}\right)
\label{eq:varying_r},
\eeq
where $r_{low}$ and $r_{high}$ are the Pearson $r$ values for the points {less than} and greater than or equal to the transition rotation period re-calculated with each transition period choice, respectively{;} $n_{low}$ is the number of stars with rotation periods below the transition, and $N$ is the total number of stars of that stellar type. {Minimizing $F_T$
of Equation (\ref{eq:varying_r})} with respect to $n_{low}$ allows us to look for a transition that maximizes the correlations, since the value of $n_{low}$ is determined by the choice of transition period. For values of $\lambda$, we still use Equations (\ref{ourlam}) and (\ref{ourerr}), but find individual values for each age regime of the dual power law.

{Stars at different points in their respective stellar activity cycles likely have different levels of spottiness for a particular age, producing a spread in the variability for stars at a given  rotation period. }To mitigate this, we  employ vertical bins of constant width {in variability of 0.1\%. This  width is based on the observed change in variability of the Sun between solar activity cycle maxima and minima from over 40 years of observations shown in \citet{Reinhold2017}, and empirical correlations between {X-ray} and bolometric variability \citep{Preminger+2011}. The general trends do not change significantly with different bin sizes.} Each bin is then represented by a single point at the mean of the data within the bin and is given a weight of the fraction of stars in that stellar type. {Trends in this binned data again do not change significantly when using the median instead of the mean.} We again quantify the statistical significance of resulting fits using the reduced chi square values from Equation (\ref{chi}).

\section{Results}

\subsection{Single Power Law}
\label{sec:onefit}

We first investigate the relationships between the square of the rotation period and the stellar variability informed by the Pearson correlation coefficient analysis. 
If the relationship were described by a power law, there would be a linear trend in log-log space that can be quantified with the Pearson correlation coefficient. We find Pearson $r$ values of $-0.55$ for M{-}type stars, $-0.58$ for K{-}type stars, $-0.40$ for G{-}type stars, and $-0.19$ for F{-}type stars. These all indicate a decreasing relationship between the two parameters, although those of the M{-}, K{-}, and G{-}type stars indicate this more strongly than that of F{-}type stars.

\begin{figure}
\begin{center}
\includegraphics[width=0.45\textwidth]{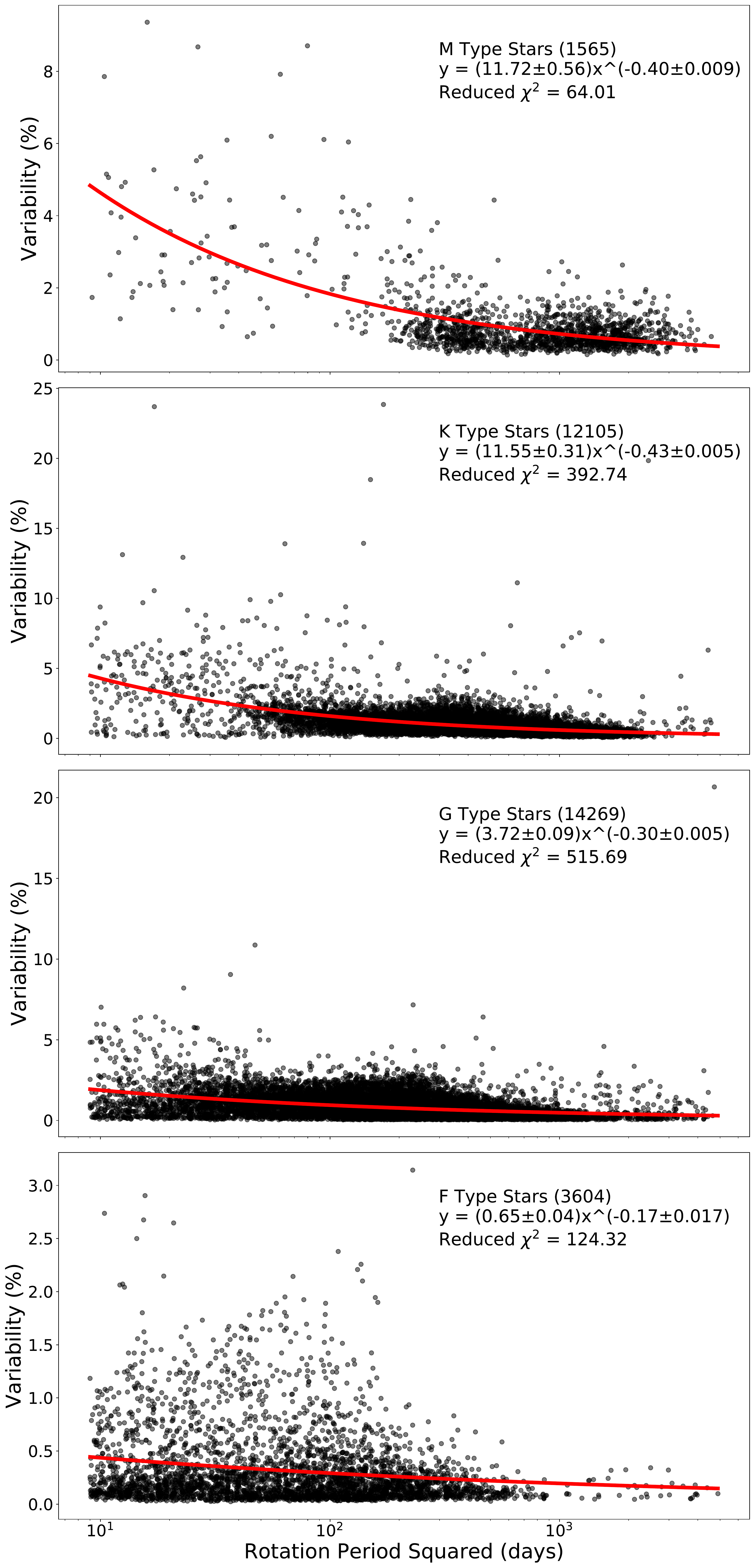}
\end{center}
\caption{Variability versus rotation period squared with a single power law fit for each stellar type. The red line on each panel represents the best fit for each stellar type. The particulars of the fits, along with the reduced $\chi^2$ values are labeled on the plot for each stellar type.}\label{fig:single_func}
\end{figure}

We characterize the decreasing variability with a single power law of the form  {$Ax^b$} for each stellar type. We use a Levenberg-Marquardt (LM) minimization routine which returns best fit parameters and errors on those derived parameters based on the gradient of the steepest descent in the minimization routine. We find the best fits (shown in Figure \ref{fig:single_func}):
{\begin{itemize}
\item[] M-type stars: $(11.72\pm0.56)x^{(-0.40\pm0.009)}$, $\chi^2=64$
\item[] K-type stars: $(11.55\pm0.31)x^{(-0.43\pm0.005)}$, $\chi^2=393$
\item[] G-type stars: $(3.72\pm0.09)x^{(-0.30\pm0.005)}$, $\chi^2=516$
\item[] F-type stars: $(0.65\pm0.04)x^{(-0.17\pm0.017)}$, $\chi^2=124$
\end{itemize}}

We find that our best fits for a single function had very large reduced $\chi^2$ values. None of these values are statistically significant, which suggests that a single power law is not the best function to represent the evolution of activity for any of these stellar types, and we therefore attempt to fit the data with two power laws.

\subsection{Dual Power Law}

\begin{figure}{
\begin{center}
\includegraphics[width=0.45\textwidth]{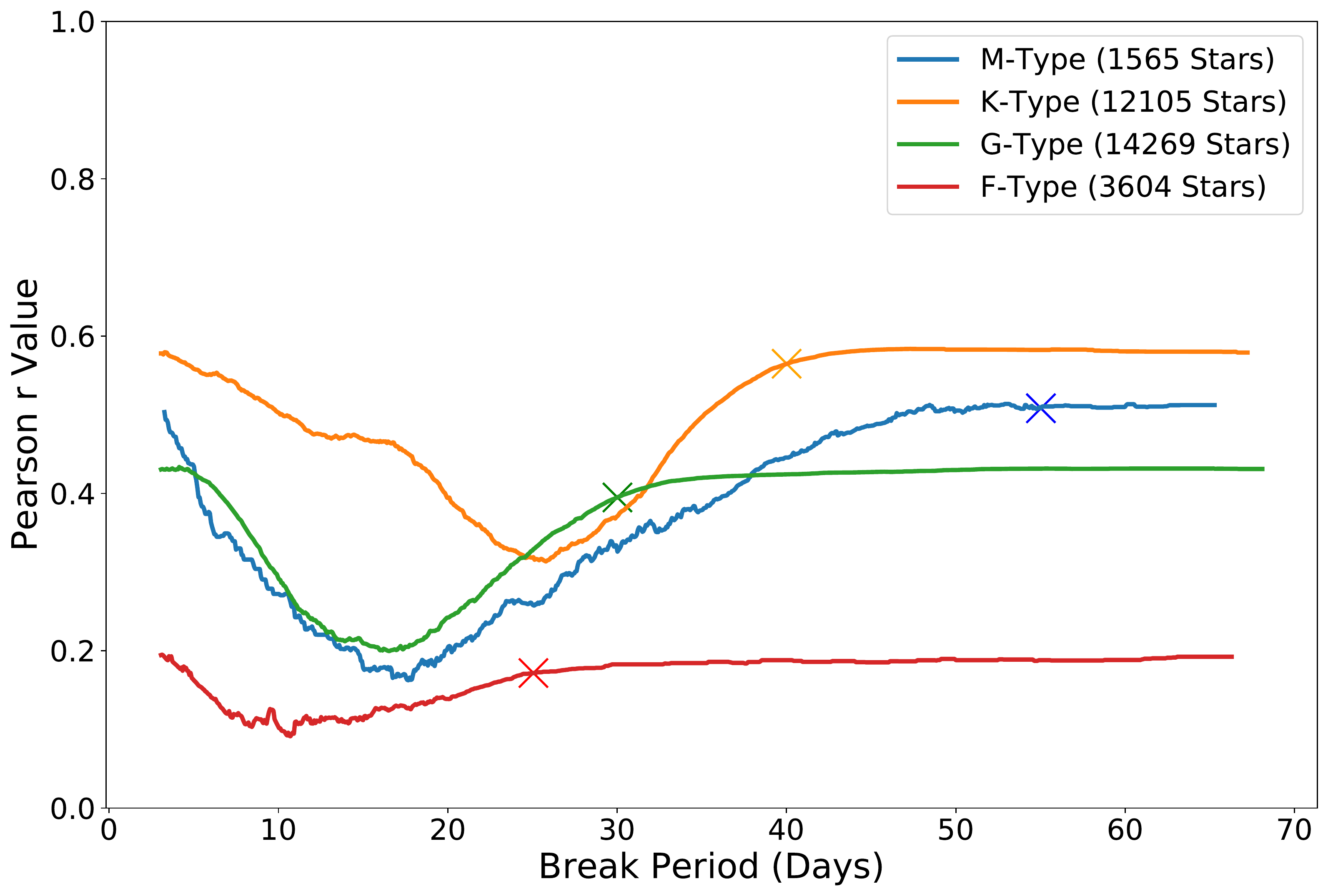}
\end{center}
\caption{Pearson $r$ values vs break period for each stellar type, blue for M-type stars, orange for K-type stars, green for G-type stars, and red for F-type stars. Each line has an x of the same color which shows the selected break period for that stellar type. The number of stars for each stellar type is shown in the upper right corner.}\label{fig:varying_pearson}}
\end{figure}

Previous work has suggested there is a change in magnetic field geometry near  the age of the Sun for Sun-like main{-}sequence stars \citep{vanSaders2016}. This, and the large reduced $\chi^2$ values from our single power law fits, prompted us to try to fit the data set with two different power laws. Rather than bias the choice of a possible power law break to the  current solar rotation period, we minimize equation (\ref{eq:varying_r}), {in order to maximize the Pearson correlation} to assess whether any of the stellar types show evidence of a dual power law when the break value is determined by the best fit. Our minimization routine for unbinned data finds no significant global minimum of the function for any of the stellar type groups so we are left to a somewhat arbitrary choice.
 Figure \ref{fig:varying_pearson} shows the weighted correlation coefficients for each stellar type along with our choice for the periods at the power law break. These {``}break periods" were simply selected as the points where the curves leveled off after the large minimum in correlation and are 55, 40, 30, and 25 days for M{-}, K{-}, G{-} and F{-}type stars, respectively. {We chose these points as they are the lowest break periods that provide the maximum weighted correlation, allowing for the highest number of stars for the fits of each power law.}

\begin{figure}
\begin{center}
\includegraphics[width=0.45\textwidth]{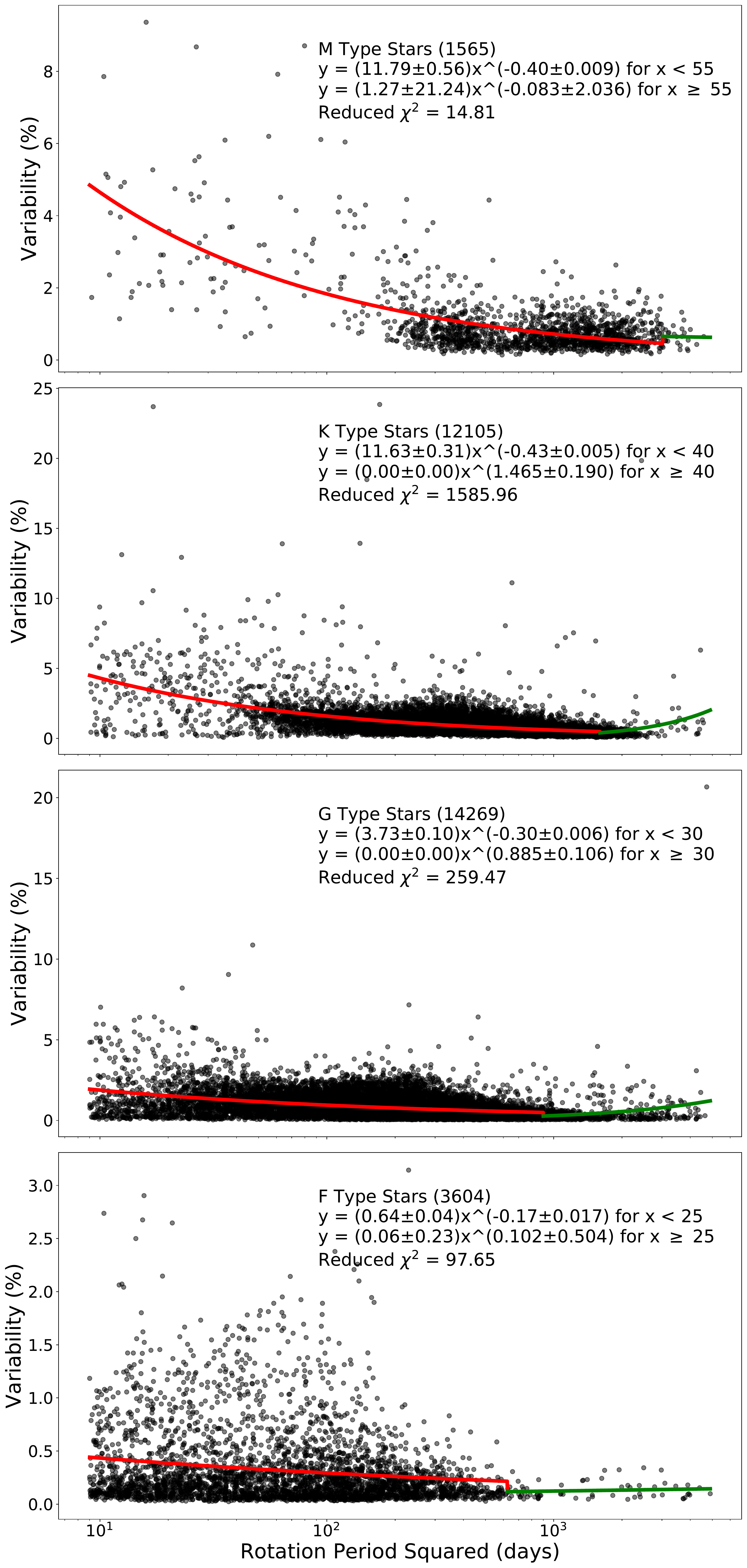}
\end{center}
\caption{Variability versus rotation period squared with dual power law fits for each stellar type split at the selected break period. The green line on each panel represents the fit for stars older than the given break period and the red line on each panel represents the fit for stars younger than the given break period. The particulars of the fits, along with the reduced $\chi^2$ values are labeled on the plot for each stellar type.}\label{fig:dual_func}
\end{figure}

Once break periods for each stellar type were  selected, we fit each with the LM minimization routine from Section \ref{sec:onefit}, but with two power laws that apply for stars with rotation periods less than or greater than the break period, respectively. We find the best fits (shown in Figure \ref{fig:dual_func}):
\begin{itemize}
\item[] {M-type stars: $(11.79\pm0.56)x^{(-0.40\pm0.009)}$ for $P_{rot} < 55$ days and $(1.27\pm21.24)x^{(-0.083\pm2.036)}$ for $P_{rot}\geq 55$ days}

\item[] {K-type stars: $(11.63\pm0.31)x^{(-0.43\pm0.005)}$ for $P_{rot} < 40$ days and $(0.00\pm0.00)x^{(1.465\pm0.190)}$ for $P_{rot} \geq 40$ days}

\item[] {G-type stars: $(3.73\pm0.10)x^{(-0.30\pm0.006)}$ for $P_{rot} < 30$ days and $(0.00\pm0.00)x^{(0.885\pm0.106)}$ for $P_{rot} \geq 30$ days}

\item[] {F-type stars: $(0.64\pm0.04)x^{(-0.17\pm0.017)}$ for $P_{rot} < 25$ days and $(0.06\pm0.23)x^{(0.102\pm0.504)}$ for $P_{rot} \geq 25$ days}
\end{itemize}

These fits produce smaller reduced $\chi^2$ values for all the stellar types with the exception of K{-}type stars: 
{14.81 for M-type stars, 1586 for K-type stars, 259.5 for G-type stars, and 97.65 for F-type stars.} Although these values are smaller than those of a single power law fit, they are still too large to be statistically significant, suggesting that other factors including each star's activity cycle may need to be considered.

\subsection{Vertical Binning}
\label{sec:vert}

Since the raw unbinned data of the previous section did not show statistical significance for a particular
power law, we considered what physical principle might justify binning the data.
Each star in the sample likely has its own stellar {activity} cycle over which the brightness will vary.
Thus all of the data points in our sample come from potentially different points in each star's activity cycle, {adding uncertainty to the exact stellar variation over rotation period squared}. To account for this spread, we introduce vertical bins into our data, each with a constant width of 0.1\% variability based on empirical correlations between {X-ray} and bolometric variability \citep{Preminger+2011} {as well as observed changes in the Sun's variability over it's known activity cycle \citep{Reinhold2017}.} We then fit the binned data with single and dual power laws as above, and assess evidence for multiple branches.  We treat the mean of each bin as a single data point with a weight given by the fraction of stars in that stellar type in the bin.

\subsubsection{Single Power Law}

 We fit our binned data with a single power law again using the LM minimization routine from Section \ref{sec:onefit}. We find the best fits:
 {\begin{itemize}
 \item[] M-type stars: $(13.91\pm2.06)x^{(-0.30\pm0.037)}$, $\chi^2=3.19$
 \item[] K-type stars: $(19.64\pm3.75)x^{(-0.31\pm0.044)}$, $\chi^2=3.71$
 \item[] G-type stars: $(15.59\pm2.62)x^{(-0.36\pm0.045)}$, $\chi^2=4.28$
 \item[] F-type stars: $(7.48\pm3.05)x^{(-0.43\pm0.113)}$, $\chi^2=1.54$
 \end{itemize}}

{These} fits of the binned data produce dramatically better reduced $\chi^2$ values for both the single and dual power law fits of the unbinned data. All of these values suggest that these fits are statistically significant.

\subsubsection{Dual Power Law}

\begin{figure}{
\begin{center}
\includegraphics[width=0.45\textwidth]{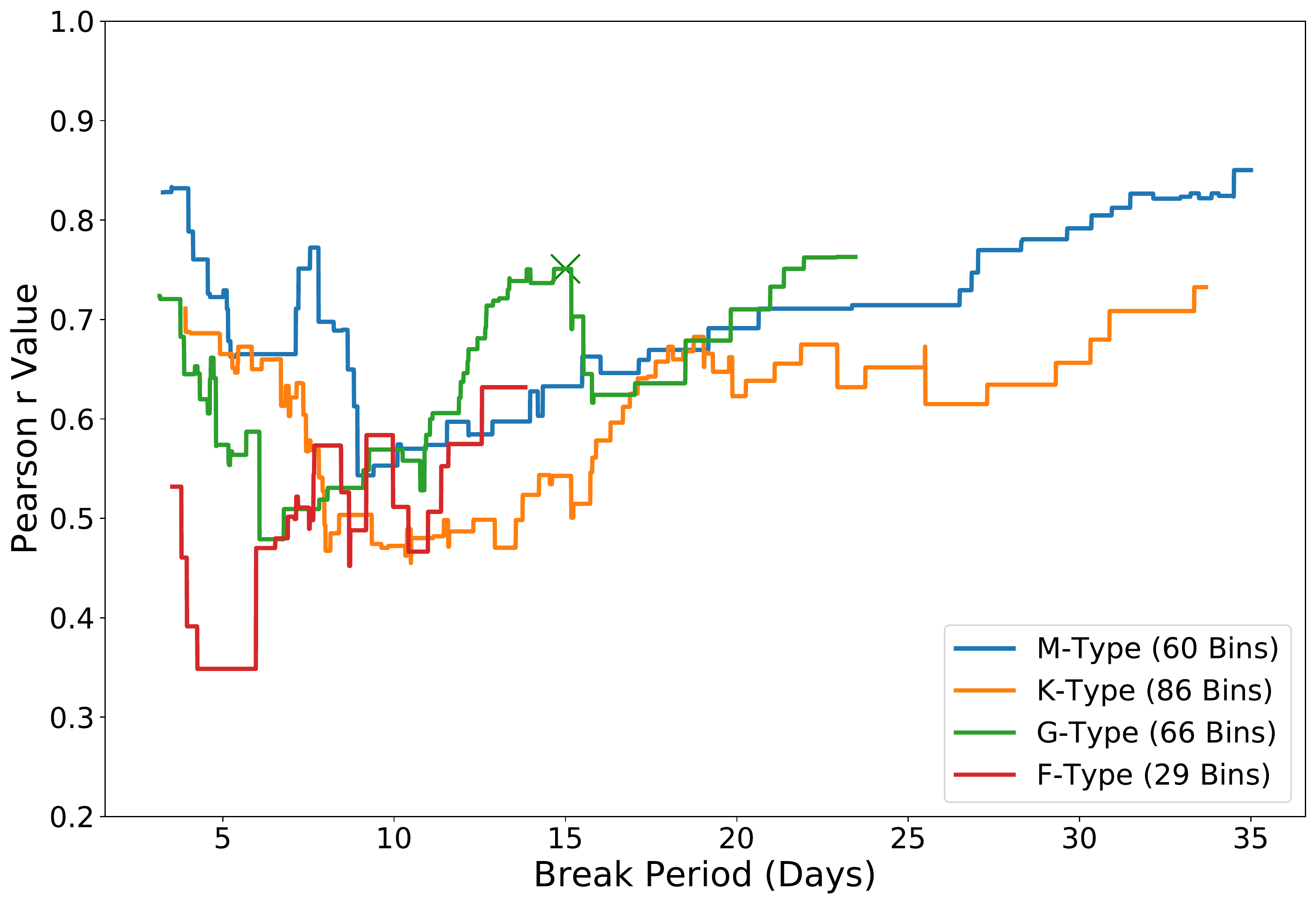}
\end{center}
\caption{Pearson $r$ values vs break period for the binned data of each stellar type, blue for M-type stars, orange for K-type stars, green for G-type stars, and red for F-type stars. As the only clear maximum in correlation occurs at 15 days for G-type stars (indicated with a green x), that is the only stellar type we fit with a dual power law. The number of bins for each stellar type is shown in the lower right corner.}\label{fig:varyingr}}
\end{figure}

To further compare our binned and unbinned results, we checked for any evidence of a dual power law fit in the binned data. We again minimize Equation (\ref{eq:varying_r}) with our binned data and again find no clear statistically significant minima that would  provide a break period choice to maximize the correlations, except possibly for G{-type} stars. Figure \ref{fig:varyingr} shows the weighted correlation coefficients for each stellar type as a function of the chosen {break} period. Only G{-}type stars show a maximum correlation that is not located at one of the edges of the possible break periods {and therefore we only attempt a dual power law fit for the binned data of G-type stars.} {The peak correlation of G-type stars} corresponds to a break period of 15 days and from this we find the fits (shown in Figure \ref{fig:vert_bins}): {$(15.11\pm2.75)x^{(-0.35\pm0.052)}$ and $(1.36\times10^{8}\pm9.79\times10^{8})x^{(-3.18\pm1.31)}$} for stars with rotation periods lower and higher than 15 days, respectively. This dual power law fit produces a lower reduced $\chi^2$ value of 1.11, suggesting a higher statistical significance of a dual power law fit than that of a single power law for the binned G{-}type stars.

\subsubsection{Branches}

\begin{figure}
\begin{center}
\includegraphics[width=0.45\textwidth]{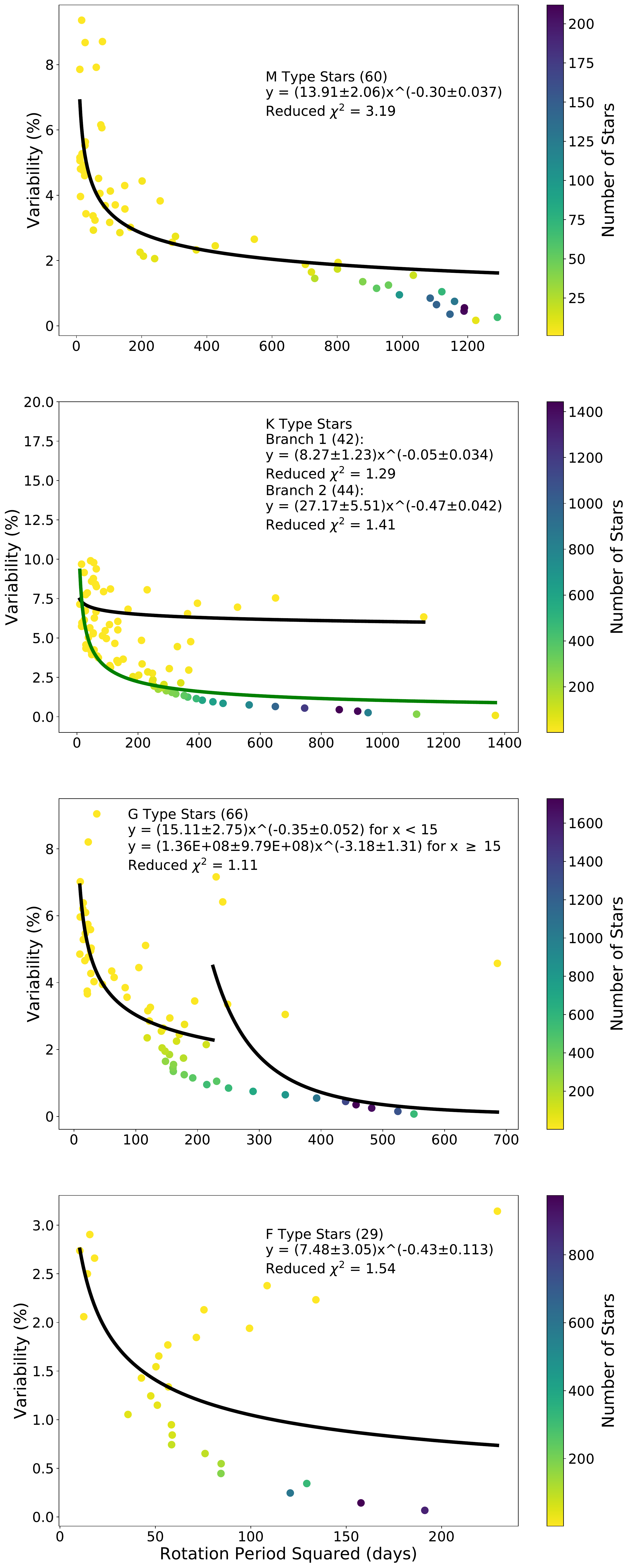}
\end{center}
\caption{Variability versus rotation period squared with power law fits for each stellar type where the sample has been split into vertical bins with a constant width of 0.1\% variability. The {black} line on the panels for M{-} and F{-}type stars represents the best fit found taking into account all bins which contain more than one star using an LM fitting routine, the black and green lines on the panel for G{-}type stars represent the fits for the bins with rotation periods less than and greater than 15 days, respectively, and the black and green lines on the panel for the K{-}type stars represent the best fits for the two branches, above and below 4.45\% variability, using an LM fitting routine. The color scale of each data point represents the number of stars in the bin represented by that point.}\label{fig:vert_bins}
\end{figure}

The scatter in the binned data suggests  another possible interpretation to test, 
namely  the existence of multiple branches in the evolution of starspot coverage with age. In particular, the binned data for K{-}type stars appears to have a structure that could be consistent with two separate branches rather than a single power law. We determine the two branches simply by grouping stars with variability above or below 4.45\%. {This division is a visual cut for where there appear to be distinct branches.} We find the best fits for K{-}type stars (shown in Figure \ref{fig:vert_bins}): {$(8.27\pm1.23)x^{(-0.05\pm0.034)}$ and $(27.17\pm5.51)x^{(-0.47\pm0.042)}$} for the stars above and below 4.45\% variability, respectively.

From these fits we find that two branches for the K{-}type stars produces smaller reduced chi square values than a single fit of the binned data, 1.24 and 1.30 for K{-}type stars with variabilities above and below 4.45\%, indeed suggesting evidence for two branches of evolution for activity of K{-}type stars.

\section{Implications for the value of $\lambda$}

Using Equations (\ref{ourlam}) and (\ref{ourerr}), we are able to determine a value for $\lambda$ for the single power law fits of our binned data, as those are the fits with reduced $\chi^2$ values indicating statistical significance. We find the values:
{\begin{itemize}
\item[] M-type stars: $0.222\pm0.030$ 
\item[] K-type stars: $0.233\pm0.035$ 
\item[] G-type stars: $0.265\pm0.036$
\item[] F-type stars: $0.316\pm0.085$
\end{itemize}}
The dual power law fit for the binned G{-}type stars also has a reduced $\chi^2$ value indicating statistical significance, and we find the $\lambda$ values 
{\begin{itemize}
\item[] $P_{rot} < 15$ days: $0.258\pm0.040$ 
\item[] $P_{rot} \geq 15$ days: $2.356\pm0.975$
\end{itemize}}
Additionally the two branches for the binned K{-}type stars have reduced $\chi^2$ values indicating statistical significance, and we find the $\lambda$ values: 
{\begin{itemize}
\item[] $V < 4.45$\% $0.034\pm0.025$
\item[] $V \geq 4.45$\% $0.350\pm0.036$
\end{itemize}}

All of the $\lambda$ values we were able to determine are within the assumed range of of $0\leq \lambda\leq \frac{1}{3}$ from \cite{Blackman2016}, with the exception of {the $\lambda$ representing G-type stars with rotation periods greater than 15 days.
The larger $\lambda$ value for this subset of the sample provides some evidence for a flattening in the functional form of the evolution of starspot coverage at a rotation period of 15 days for G-type stars. This period is different from that associated with a previously proposed weakening of magnetic braking at the solar age, corresponding to the 24.5 day rotation period \citep{vanSaders2016,Metcalfe+2017}. A weakening of magnetic braking through some kind of dynamo transition to higher multiploles could be related to the increase in $\lambda$: a higher $\lambda$ means more sensitivity of starspot coverage to x-ray luminosity. In turn, a reduction in magnetic braking for a given average magnetic field strength implies a a higher fraction of smaller scale magnetic structures. Since it is these smaller structures which produce starspots, we can speculate that $\lambda$ could increase with a reduction in open field lines. In any case, more observational and theoretical work is needed for the ``first principles" understanding of $\lambda$ from stellar dynamo theory, but the empirically determined value is of practical use for simple holistic modeling.}

\section{Conclusions}

To investigate the {starspot} coverage of main{-}sequence stars, we have used a sample of over 30,000 {\textit{Kepler}} stars with previously measured activity and rotation periods covering M, G, K, and F spectral types. Using the Pearson correlation coefficient, we determined that there exists a decrease in {starspot} coverage with increasing {rotation period---a proxy for age---}for each of these spectral types. We tried to describe this relationship in the unbinned raw data with single and dual power law fits using a LM minimization routine, and {found} high reduced $\chi^2$ values for each stellar type in both cases, although those of the dual power law fits are somewhat smaller than those of the single power law for all types except for the K{-}type stars.

After introducing vertical bins to account for the variability within {activity} cycle periods of each individual star, we {found much lower reduced $\chi^2$ values than those of the unbinned data, indicating} statistically significant fits for each stellar type. We found no evidence for power law breaks in the binned data except for the G{-}type stars, in which a break occurs at a rotation period of 15 days. {This} dual power law fit for G-type stars produced a lower reduced $\chi^2$ value than the single power law. We also found evidence of two separate branches for the evolution of K-type stars, those above and below 4.45\% variability, which further decreases the reduced $\chi^2$ values, indicating a higher statistical significance for these branches than a single power law.  {We compared 
fits using bin means and medians for all stellar types, and found the only significant difference to be somewhat reduced evidence for the two branches for the K-type stars.}

{From our fits, we determined values for the model parameter $\lambda$ used in \cite{Blackman2015} to account for for the decreasing evolution of starspot coverage with age. We found values in concordance with their assumed range of $0\leq\lambda\leq\frac{1}{3}$ for the single power law fit for all stellar types. We also found concordance for both branches of K-type stars. For the dual power law fit to G-type stars, we found concordance at rotation periods below 15 days. The $\lambda$ value for G-type stars with rotation periods greater than 15 days is over $2\sigma$ outside the range of \cite{Blackman2015}, indicating some evidence for a flattening in the functional form of the evolution of starspot coverage. This piques further interest into a potential transition for G-type stars.}

{Although vertical binning was a practical step toward alleviating the influence of variability within each stars stellar cycle on the spread in the raw data, future work could improve upon this if large samples were to ever become available with known activity cycles, so that stars can be compared at the same phase in their cycles. Additionally, more observational and theoretical work is needed to understand the parameter $\lambda$ from stellar dynamo theory.}

\section*{Acknowledgments} 
We thank  the referee for a helpful, meticulous report.
 EB acknowledges support from NSF Grant  AST-1813298, KITP (UC Santa Barbara) funded by  NSF Grant PHY-1748958, and Aspen Center for Physics funded by NSF Grant PHY-1607611.

\bibliographystyle{mn2e}
\bibliography{spotsbib}

\begin{thebibliography}{49}
\expandafter\ifx\csname natexlab\endcsname\relax\def\natexlab#1{#1}\fi

\bibitem[{{Baliunas} {et~al}\mbox{.}(1995){Baliunas}, {Donahue}, {Soon},
  {Horne}, {Frazer}, {Woodard-Eklund}, {Bradford}, {Rao}, {Wilson}, {Zhang},
  {Bennett}, {Briggs}, {Carroll}, {Duncan}, {Figueroa}, {Lanning}, {Misch},
  {Mueller}, {Noyes}, {Poppe}, {Porter}, {Robinson}, {Russell}, {Shelton},
  {Soyumer}, {Vaughan}, \& {Whitney}}]{Baliunas1995}
{Baliunas} S.~L. {et~al.}, 1995, \apj, 438, 269

\bibitem[{{Baliunas} \& {Vaughan}(1985)}]{Baliunas1985}
{Baliunas} S.~L., {Vaughan} A.~H., 1985, Annual Review of Astronomy and
  Astrophysics, 23, 379

\bibitem[{{Blackman} \& {Owen}(2016)}]{Blackman2016}
{Blackman} E.~G., {Owen} J.~E., 2016, \mnras, 458, 1548

\bibitem[{{Blackman} \& {Thomas}(2015)}]{Blackman2015}
{Blackman} E.~G., {Thomas} J.~H., 2015, \mnras, 446, L51

\bibitem[{{B{\"o}hm-Vitense}(2007)}]{Bohm2007}
{B{\"o}hm-Vitense} E., 2007, \apj, 657, 486

\bibitem[{{Brandenburg}, {Saar} \& {Turpin}(1998){Brandenburg}, {Saar}, \&
  {Turpin}}]{Brandenburg1998b}
{Brandenburg} A., {Saar} S.~H., {Turpin} C.~R., 1998, \apj, 498, L51

\bibitem[{{Charbonneau}(2014)}]{Charbonneau2014}
{Charbonneau} P., 2014, \araa, 52, 251

\bibitem[{{Cranmer} \& {Saar}(2011)}]{Cranmer2011}
{Cranmer} S.~R., {Saar} S.~H., 2011, \apj, 741, 54

\bibitem[{{Dmitrienko} \& {Savanov}(2017)}]{Dmitrienko2017}
{Dmitrienko} E.~S., {Savanov} I.~S., 2017, Astronomy Reports, 61, 122

\bibitem[{{do Nascimento}, {Saar} \& {Anthony}(2015){do Nascimento}, {Saar}, \&
  {Anthony}}]{doNascimento2015}
{do Nascimento}, Jr. J.-D., {Saar} S.~H., {Anthony} F., 2015, in Cambridge
  Workshop on Cool Stars, Stellar Systems, and the Sun, Vol.~18, 18th Cambridge
  Workshop on Cool Stars, Stellar Systems, and the Sun

\bibitem[{{Domingo} {et~al}\mbox{.}(2009){Domingo}, {Ermolli}, {Fox},
  {Fr{\"o}hlich}, {Haberreiter}, {Krivova}, {Kopp}, {Schmutz}, {Solanki},
  {Spruit}, {Unruh}, \& {V{\"o}gler}}]{Domingo2009}
{Domingo} V. {et~al.}, 2009, \ssr, 145, 337

\bibitem[{{Dressing} \& {Charbonneau}(2013)}]{Dressing2013}
{Dressing} C.~D., {Charbonneau} D., 2013, \apj, 767, 95

\bibitem[{{Drilling} \& {Landolt}(2000)}]{Drilling2000}
{Drilling} J.~S., {Landolt} A.~U., 2000, {Normal Stars}, {Cox} A.~N., ed., p.
  381

\bibitem[{{Duvall} {et~al}\mbox{.}(1984){Duvall}, {Dziembowski}, {Goode},
  {Gough}, {Harvey}, \& {Leibacher}}]{Duvall1984}
{Duvall}, T.~L. J., {Dziembowski} W.~A., {Goode} P.~R., {Gough} D.~O., {Harvey}
  J.~W., {Leibacher} J.~W., 1984, \nat, 310, 22

\bibitem[{{Hartmann} \& {Noyes}(1987)}]{Hartmann1987}
{Hartmann} L.~W., {Noyes} R.~W., 1987, \araa, 25, 271

\bibitem[{{Kraft}(1967)}]{Kraft1967}
{Kraft} R.~P., 1967, \apj, 150, 551

\bibitem[{{Mamajek} \& {Hillenbrand}(2008)}]{Mamajek+2008}
{Mamajek} E.~E., {Hillenbrand} L.~A., 2008, ApJj, 687, 1264

\bibitem[{{McQuillan}, {Aigrain} \& {Mazeh}(2013){McQuillan}, {Aigrain}, \&
  {Mazeh}}]{McQuillan2013}
{McQuillan} A., {Aigrain} S., {Mazeh} T., 2013, \mnras, 432, 1203

\bibitem[{{McQuillan}, {Mazeh} \& {Aigrain}(2014){McQuillan}, {Mazeh}, \&
  {Aigrain}}]{McQuillan2014}
{McQuillan} A., {Mazeh} T., {Aigrain} S., 2014, \apjs, 211, 24

\bibitem[{{Messina} \& {Guinan}(2002)}]{Messina2002}
{Messina} S., {Guinan} E.~F., 2002, \aap, 393, 225

\bibitem[{{Metcalfe} \& {van Saders}(2017)}]{Metcalfe+2017}
{Metcalfe} T.~S., {van Saders} J., 2017, Solar Physics, 292, 126

\bibitem[{{Micela} {et~al}\mbox{.}(1985){Micela}, {Sciortino}, {Serio},
  {Vaiana}, {Bookbinder}, {Golub}, {Harnden}, \& {Rosner}}]{Micela1985}
{Micela} G., {Sciortino} S., {Serio} S., {Vaiana} G.~S., {Bookbinder} J.,
  {Golub} L., {Harnden}, Jr. F.~R., {Rosner} R., 1985, \apj, 292, 172

\bibitem[{{Montesinos} {et~al}\mbox{.}(2001){Montesinos}, {Thomas}, {Ventura},
  \& {Mazzitelli}}]{Montesinos2001}
{Montesinos} B., {Thomas} J.~H., {Ventura} P., {Mazzitelli} I., 2001, \mnras,
  326, 877

\bibitem[{{Notsu} {et~al}\mbox{.}(2019){Notsu}, {Maehara}, {Honda}, {Hawley},
  {Davenport}, {Namekata}, {Notsu}, {Ikuta}, {Nogami}, \&
  {Shibata}}]{Notsu2019}
{Notsu} Y. {et~al.}, 2019, arXiv e-prints, arXiv:1904.00142

\bibitem[{{Noyes} {et~al}\mbox{.}(1984){Noyes}, {Hartmann}, {Baliunas},
  {Duncan}, \& {Vaughan}}]{Noyes1984a}
{Noyes} R.~W., {Hartmann} L.~W., {Baliunas} S.~L., {Duncan} D.~K., {Vaughan}
  A.~H., 1984, \apj, 279, 763

\bibitem[{{Noyes}, {Weiss} \& {Vaughan}(1984){Noyes}, {Weiss}, \&
  {Vaughan}}]{Noyes1984b}
{Noyes} R.~W., {Weiss} N.~O., {Vaughan} A.~H., 1984, \apj, 287, 769

\bibitem[{{Ol{\'a}h}, {Koll{\'a}th} \& {Strassmeier}(2000){Ol{\'a}h},
  {Koll{\'a}th}, \& {Strassmeier}}]{Olah2000}
{Ol{\'a}h} K., {Koll{\'a}th} Z., {Strassmeier} K.~G., 2000, \aap, 356, 643

\bibitem[{{Ol{\'a}h} \& {Strassmeier}(2002)}]{Olah2002}
{Ol{\'a}h} K., {Strassmeier} K.~G., 2002, Astronomische Nachrichten, 323, 361

\bibitem[{{Pallavicini} {et~al}\mbox{.}(1981){Pallavicini}, {Golub}, {Rosner},
  {Vaiana}, {Ayres}, \& {Linsky}}]{Pallavicini1981}
{Pallavicini} R., {Golub} L., {Rosner} R., {Vaiana} G.~S., {Ayres} T., {Linsky}
  J.~L., 1981, \apj, 248, 279

\bibitem[{{Parker}(1955)}]{Parker1955}
{Parker} E.~N., 1955, \apj, 122, 293

\bibitem[{{Petrovay} \& {van Driel-Gesztelyi}(1997)}]{Petrovay1997}
{Petrovay} K., {van Driel-Gesztelyi} L., 1997, \solphys, 176, 249

\bibitem[{{Pizzolato} {et~al}\mbox{.}(2003){Pizzolato}, {Maggio}, {Micela},
  {Sciortino}, \& {Ventura}}]{Pizzolato2003}
{Pizzolato} N., {Maggio} A., {Micela} G., {Sciortino} S., {Ventura} P., 2003,
  \aap, 397, 147

\bibitem[{{Preminger}, {Chapman} \& {Cookson}(2011){Preminger}, {Chapman}, \&
  {Cookson}}]{Preminger+2011}
{Preminger} D.~G., {Chapman} G.~A., {Cookson} A.~M., 2011, \apjl, 739, L45

\bibitem[{{Randich}(2000)}]{Randich2000}
{Randich} S., 2000, in Astronomical Society of the Pacific Conference Series,
  Vol. 198, Stellar Clusters and Associations: Convection, Rotation, and
  Dynamos, {Pallavicini} R., {Micela} G., {Sciortino} S., eds., p. 401

\bibitem[{{Reiners}, {Sch{\"u}ssler} \& {Passegger}(2014){Reiners},
  {Sch{\"u}ssler}, \& {Passegger}}]{Reiners2014}
{Reiners} A., {Sch{\"u}ssler} M., {Passegger} V.~M., 2014, \apj, 794, 144

\bibitem[{{Reinhold}, {Cameron} \& {Gizon}(2017){Reinhold}, {Cameron}, \&
  {Gizon}}]{Reinhold2017}
{Reinhold} T., {Cameron} R.~H., {Gizon} L., 2017, \aap, 603

\bibitem[{{Saar} \& {Brandenburg}(1999)}]{Saar1999}
{Saar} S.~H., {Brandenburg} A., 1999, \apj, 524, 295

\bibitem[{{Savanov} \& {Dmitrienko}(2017{\natexlab{a}})}]{Savanov2017b}
{Savanov} I.~S., {Dmitrienko} E.~S., 2017{\natexlab{a}}, Astronomy Reports, 61,
  996

\bibitem[{{Savanov} \& {Dmitrienko}(2017{\natexlab{b}})}]{Savanov2017a}
{Savanov} I.~S., {Dmitrienko} E.~S., 2017{\natexlab{b}}, Astronomy Reports, 61,
  461

\bibitem[{{Savanov} \& {Dmitrienko}(2018)}]{Savanov2018}
{Savanov} I.~S., {Dmitrienko} E.~S., 2018, Astronomy Reports, 62, 238

\bibitem[{{Skumanich}(1972)}]{Skumanich1972}
{Skumanich} A., 1972, \apj, 171, 565

\bibitem[{{van Saders} {et~al}\mbox{.}(2016){van Saders}, {Ceillier},
  {Metcalfe}, {Silva Aguirre}, {Pinsonneault}, {Garc{\'{\i}}a}, {Mathur}, \&
  {Davies}}]{vanSaders2016}
{van Saders} J.~L., {Ceillier} T., {Metcalfe} T.~S., {Silva Aguirre} V.,
  {Pinsonneault} M.~H., {Garc{\'{\i}}a} R.~A., {Mathur} S., {Davies} G.~R.,
  2016, Nature, 529, 181

\bibitem[{{Vaughan} \& {Preston}(1980)}]{Vaughan1980}
{Vaughan} A.~H., {Preston} G.~W., 1980, Publications of the Astronomical
  Society of the Pacific, 92, 385

\bibitem[{{Vaughan}, {Preston} \& {Wilson}(1978){Vaughan}, {Preston}, \&
  {Wilson}}]{Vaughan1978}
{Vaughan} A.~H., {Preston} G.~W., {Wilson} O.~C., 1978, Publications of the
  Astronomical Society of the Pacific, 90, 267

\bibitem[{{Vidotto} {et~al}\mbox{.}(2014){Vidotto}, {Gregory}, {Jardine},
  {Donati}, {Petit}, {Morin}, {Folsom}, {Bouvier}, {Cameron}, {Hussain},
  {Marsden}, {Waite}, {Fares}, {Jeffers}, \& {do Nascimento}}]{Vidotto2014}
{Vidotto} A.~A. {et~al.}, 2014, \mnras, 441, 2361

\bibitem[{{Vilhu}(1984)}]{Vilhu1984}
{Vilhu} O., 1984, \aap, 133, 117

\bibitem[{{Wilson}(1966)}]{Wilson1966}
{Wilson} O.~C., 1966, \apj, 144, 695

\bibitem[{{Wilson}(1978)}]{Wilson1978}
{Wilson} O.~C., 1978, \apj, 226, 379

\bibitem[{{Wright} {et~al}\mbox{.}(2011){Wright}, {Drake}, {Mamajek}, \&
  {Henry}}]{Wright2011}
{Wright} N.~J., {Drake} J.~J., {Mamajek} E.~E., {Henry} G.~W., 2011, \apj, 743,
  48

\end{thebibliography}

\appendix
\section{Comparison of Medians and Means of Binned Data}

\begin{figure*}{
\begin{center}
\includegraphics[width=0.85\textwidth]{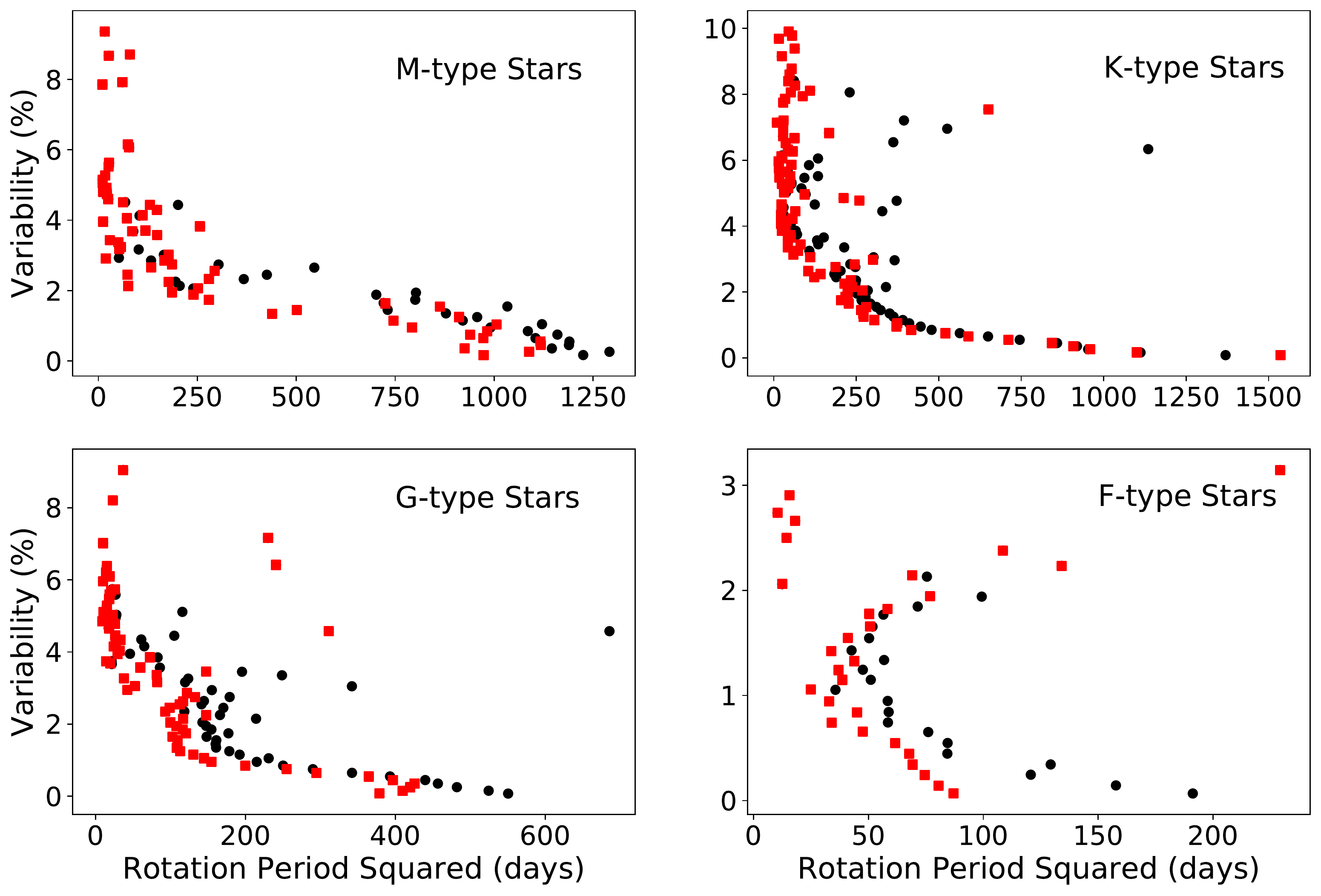}
\end{center}
\caption{Comparisons of trends seen when using the medians vs the means of the rotation period squared and variability of the binned data for each stellar type. The black circles in each panel represent the means of the bins and the red squares represent the medians of the bins. In each case the trends seen in the medians and means are the same with the exception of more evidence for a second branch in the means of K-type stars than in the medians.}\label{fig:med_vs_mean}}
\end{figure*}

{To ensure that the trends in the binned data are not affected by the use of the mean values, we compare the trends seen in the medians of the bins with those seen in means of the bins as shown in Figure \ref{fig:med_vs_mean}. We use the same bins in variability of constant width 0.1\% as in Section \ref{sec:vert} and represent each bin with either the average or the median rotation period and variability.  As shown in  in Appendix A, the same trends are seen in the medians and means for M-, G-, and F-type stars. For K-type stars there appears to be more evidence for a second branch in the means of the bins than in the medians, but there does still exist some evidence for a second branch in the medians.}

\label{lastpage}

\end{document}